**Understanding and enhancing the impact-induced tension of a falling chain**

J. Pantaleone*, Department of Physics, University of Alaska Anchorage, Anchorage Alaska, 99508

When a falling chain strikes a surface, it can accelerate downwards faster than free-fall. This counterintuitive effect occurs when a tension is created in the chain above where it strikes the surface. The size of this tension, and how it is produced, depend on the type of chain used. For a chain made of rods that are slightly tilted from horizontal, the impact-induced tension is readily observable. Here are reported experimental observations on such a falling chain for two different situations: when the chain strikes an inclined surface, and when the chain's mass density decreases with height. It is found that both of these arrangements can increase the downward acceleration. To quantitatively describe these observations a theoretical model is developed. The model successfully predicts the chain's position and velocity, even when the top end approaches the surface, without any free parameters. The model also predicts that uniform rods are practically the best for producing large tensions.

## I. INTRODUCTION

Moving strings and chains have been a subject of study for over 2 centuries[1-27]. This is primarily because these systems are essential elements of mechanical technology, used in bicycles, fishing poles, space tethers and more. Moving strings and chains are also useful for educational purposes, as relatively simple examples for illustrating the dynamics of continuous mass systems. Of special interest are situations where the energy of the moving string or chain becomes concentrated in a small region, magnifying the speed in that region. Well-known examples of this are the crack of a whip[5,8,9] or the fall of a folded chain.[3,4,6,7,10,25] In the last few years, two new motion-enhancing systems have been discovered, occurring when a chain is lifted off of or laid down onto a surface. The former is responsible for the chain fountain,[14-21,23,24] where a chain leaps out of a beaker as it falls to the ground. The latter, chain lay-down onto a surface, is the subject of this paper.

When a chain falls onto a surface the downward motion can be faster than free-fall---the chain is pulled towards the surface.[11-13,22,23] The forces acting on the falling part of the chain are sketched in Fig. 1. Applying Newton's second law to this system gives an equation of motion for the falling part of the chain that can be written as

$$M \frac{d^2 z}{dt^2} = -Mg - F. \tag{1}$$

Here $M$ is the mass of the chain inside the dashed region; $g$ is the acceleration of gravity; and $F$ is the force acting at the bottom of the dashed region. The position $z$ has been chosen here to be the height of the chain; however all positions on the chain inside the dashed region have the same velocity and acceleration. The force $F$ is the tension created when the chain strikes the surface. Because $F$ acts in the same direction as the velocity, the speed of the falling chain is increased. Eq. (1) treats the chain as a smooth, continuous system, and so it applies when the internal structure of the chain (e.g. the links) can be averaged over. Eq. (1) can be derived in many different ways: applying

Newton's second law to each link in the dashed region and then summing all of these equations,[23] or by using Euler-Lagrange methods,[27] or by using the formalism for variable mass systems given in most introductory physics textbooks.[28,29] In this latter formalism, the relative velocity of the ejected mass (the mass leaving the dashed region) with respect to the central mass (the mass in the dashed region) is zero, so the "momentum flux" term proportional to the rate of mass change vanishes, giving Eq. (1). This falling chain system is a nice example for demonstrating that, for variable mass systems, it is incorrect to simply replace the mass times the acceleration with the derivative of the momentum with respect to time,[28,29] since doing so here would give an additional term in Eq. (1) and then the falling chain would not be in free-fall when the tension, $F$, vanishes. Most introductory physics textbooks that discuss variable mass systems use the example of a rocket; however, the falling chain has the advantage that it can be readily realized in an introductory physics lab.

A form for the force $F$ can be deduced from dimensional analysis. Assuming that the impact point is far from the top end of the chain, and that we are only interested in the large scale behavior of the chain (so the structure of the chain can be averaged over), then only the averaged variables local to the impact point are relevant: the velocity, $dz/dt$, and the linear mass density, $\lambda$. The only combination of these variables with the correct units is

$$F = \alpha \lambda \left(\frac{dz}{dt}\right)^2 \tag{2}$$

where $\alpha$ is a dimensionless constant. Eq. (2) has been successfully used to quantitatively describe the experimental observations of impact-induced tension[11,12,22,26] and also the chain fountain.[18,21] In some of these references somewhat different notation is used,[30] but all of the explanations are equivalent to the parameterization in Eq. (2). The value of $\alpha$ must be determined by other considerations.

The impact-induced tension was first experimentally observed for a falling bead chain (also called a ball chain) by Hamm and Geminard.[11] A more recent experiment with a bead chain measured the value $\alpha \approx 0.12$ when falling onto a horizontal surface (and only when far from the top and bottom ends of the chain), and smaller values when the surface was inclined.[22] The impact-induced tension was also observed for a falling chain made of rods slightly tilted from the horizontal;[12] however, an experimental value for $\alpha$ was not reported there. Observations of a falling chain made of open metal links did not exhibit a sizeable acceleration increase,[12,13] with $\alpha \approx 0.0$. These experiments demonstrate that the size of $\alpha$ depends on the internal structure of the chain and on how the chain interacts with the surface.

An upper limit on the value of $\alpha$ can be found by assuming that the energy of the chain is conserved. In that case, when a length $\Delta z$ of the chain comes to rest on the surface, all of its kinetic energy, $\Delta K = \frac{1}{2} \lambda \Delta z (dz/dt)^2$, is assumed to be transferred to the chain above. The energy transfer occurs because the impact-induced tension $F$ does work $\Delta W = F \Delta z$ on the chain above. Thus energy conservation corresponds to $\Delta K = \Delta W$, which yields the value $\alpha = \frac{1}{2}$, in agreement with Refs. 12 and 27. This value is substantially larger than the values measured for chains falling onto a surface. Thus, while energy conservation sometimes does a reasonable job of describing smoothly unfolding chains (see e.g. Refs. 6, 10 and references therein), it does not work very well for describing chains being laid down onto (or being lifted off of[18,21]) a surface.

For a bead chain, there is a theoretical model of how the tension force is created when the chain is laid down onto a surface.[26] According to this model, when a bead of the chain collides with the surface it compresses the bottommost link and also pushes the bead above the collision point transversely. This transverse motion creates a rotation of the link above the collision point about the chain above it. It is this swinging, pendular motion of the link above the collision point that creates the tension for a bead chain.

Here we examine the dynamics of the other type of chain observed to exhibit an impact-induced tension, the one made from rods with a small tilt from horizontal. This type of chain was devised by Grewal, Johnson and Ruina[12] for demonstrating the existence of the impact-induced tension because the mechanism creating the tension force is particularly easy to understand conceptually. When a rod is struck on one end it rotates such that the other end will move opposite to the striking force, producing a downward impulse on the chain above it. This "impulse" mechanism is quite different than the "pendular" mechanism of the bead chain model, since the impulse model is relatively instantaneous while the pendular model is sensitive to how much rotation occurs between the collision of one link and the next. The rod chain is chosen for study here because (1) the mechanism for producing the tension has not previously been studied quantitatively, (2) the size of $\alpha$ should be relatively large, and (3) the calculation of the tension from basic physical principles should be relatively straightforward.

The results of two experiments are presented in section II. Each experiment is designed to further enhance the motion of the falling chain and also to test the comparison of the falling chain with the theoretical model. One experiment involves a falling chain with a mass density that decreases with height, and the other involves falling chains with uniform density striking an inclined surface. Each of these experiments is interesting, but for different reasons.

For the chain with a mass density that decreases with height, the effects of the impact-induced tension will be magnified over that of a uniform mass density chain. This is because there is a smaller mass above the impact point, so larger accelerations and displacements will be produced. This enhancement of the motion is somewhat similar to what occurs in other dynamical systems with decreasing mass density, such as a whip[5,8,9] or a Galilean cannon (also known as an Astroblaster).

In the other set of experiments, the effects of the motion enhancement were measured as a function of the inclination of the surface. Tilting the surface also has the potential to increase motion enhancement. This is because at large inclinations the rods do not accumulate in a pile on the surface but instead slide down the surface. Thus the falling chain does not strike a loose pile of rods, but instead each falling rod strikes a hard, smooth surface. Besides enhancing the motion, experimental observations on an inclined surface are interesting for several other reasons: a collision between a rod and a single smooth surface is easier to describe theoretically; the collision process is fully three dimensional in the motion of the rod; and these measurements can be compared to recent similar measurements for falling bead chains.[22]

A theoretical explanation of these experimental results is presented in section III, where the tension is calculated for rod lay-down under a variety of conditions. Section III.A considers the theoretically simplest case, a link of an infinite chain striking a smooth, horizontal surface.[31] Section III.B considers the more realistic case of a chain of finite length striking a horizontal surface, and determines the modification to Eq. (2) that is necessary to describe what happens as the chain end approaches the surface. Section III.C considers the case of an infinite chain striking an inclined surface, which is useful because no pile forms for highly inclined surfaces. Section III.D discusses how to combine the results of the previous sections to get the impact-induced tension for a finite chain striking an inclined surface. This section also calculates the predictions of the theoretical model and compares them with the current experimental results. Section IV discusses the implications of these results in the context of other falling chain experiments.

## II. EXPERIMENTAL OBSERVATIONS

The effects of the impact-induced tension were quantitatively measured by observing the relative motion of two similar chains released together. One of the falling chains was in free-fall while the other chain struck a surface. The small tension created from striking a surface increases the downward acceleration of the upper part of that chain, causing a vertical separation in the position of the top rods of the two chains. This vertical separation is an unambiguous indicator of the presence of the impact-induced tension that is relatively insensitive to many experimental uncertainties.[12]

The falling chains were observed with a video camera. The camera used was a Sony model NEX-FS700R operating at 480 fps with 1920 by 1080 resolution. The videos were analyzed using the free video analysis software *Tracker*.[32] This position data was used to measuring the vertical separation of the two chains and also to calculate the speed of the falling chains. These speed measurements are inherently more uncertain, and a systematic uncertainty for the speed measurements has been included where indicated in the results presented below.

Important for the success of these experiments is a mechanism that releases the two chains as simultaneously as possible. Past experiments used different types of mechanisms to do this. One group used a spring driven system,[12] and another group used an electromagnetic system.[22] After building a couple different release mechanisms, a gravity driven mechanism was used for the experiments presented here.[33] This mechanism consists of two parallel metal rods connected to a wooden board. At the bottom rear of this board are two hinges, which connect the board to a stationary frame mounted to the wall. The two rods and the board move together, with the rods able to rotate freely from a horizontal orientation to down to past vertical. When setting up the device, the rods are placed in the horizontal position and kept from rotating by a string attached to each rod and going over a wooden peg above the rods. When the peg is retracted, the rods are released and they swing down and away from the chains. The mass distribution in the board-rods system is such that the initial downward linear acceleration at the end of the rods is greater than *g*, so that the release of the chains is smooth. This chain release mechanism has many advantages. The mechanism is easy to set-up and it does not require an electric current to hold the chains in place. Also it is easy to release by simply pulling a string to retract the peg, thus a single person can do the experiment. Most importantly, variations in the relative release time between the two chains were too small to detect using the gravity driven release mechanism.

The chains used in these experiments were built of wooden dowels tilted slightly from horizontal following the general design used in Ref. 12; see Fig. 2. All of the chains had approximately the same length, 116.5 +/- 0.3 cm from the middle of the top rod to the middle of the bottom rod. All of the chains used dowels that were 10.2 cm long. The dowels were connected together using 130 lb test fishing "superline." This line is designed to have minimal stretch and the chains did not appreciably lengthen when hung vertically under their own weight. The dowels and fishing line were connected together using a cyanoacrylate adhesive.

For the experiments where the inclination was varied, the chains with a uniform mass distribution were used. They were made using 25 poplar dowels in each chain, where the dowel diameters had a manufacturer's stated value of 0.5 in (1.27 cm). The rods were tilted by 12° (on average) from horizontal; see Fig. 2 top. The chains with a mass density decreasing with height were made using 20 poplar dowels with 4 dowels each of the manufacturer's stated diameters of 1.25 in (3.2 cm), 1 in (2.5 cm), 0.75 in (1.9 cm), 0.5 in (1.3 cm) and 0.25 in (0.63 cm). These 20 dowels were distributed along the chain with a uniform separation between the outer edges of the dowels and had an average tilt angle of 21°; see Fig. 2 bottom.

For all experiments the surface was placed a distance of 18 cm below the middle of the bottom dowel. Larger fall distances are predicted to give slightly larger effects, but these were not used due to constraints from the space available for doing the experiments and from the speed of the video camera. This 18 cm fall distance was kept consistent for different runs by using a plumb bob of fixed length to measure from the end of the support rod on the release mechanism to the surface. However the location where a falling rod first made contact was typically somewhat above this point on the surface. This occurred for several reasons: at shallow inclinations it was because a pile of rods would accumulate, and at steep inclinations it was because of the width of the dowel and also because of some small, transverse rotations of the rods.

The surface the chain struck was 1.9 cm thick plywood with a thermally fused melamine laminate surface (obtained from a local hardware store where it is sold for use as shelving). The rigid body interaction parameters between a wooden dowel and this surface were directly measured. The coefficient of static friction was found to be $\mu \approx 0.2$, as determined by measuring the inclination angle at which a dowel at rest on the surface started to slide lengthwise. No difference was discernable between the kinetic and static coefficients of friction. To determine a coefficient of restitution, the collision between the bottom rod of the uniform density chain with the horizontal surface was videotaped with the camera as close as possible to the collision point. It was found that during the collision, the change in velocity of the upper end of the dowel was too small to measure while the lower end rebounded with the coefficient of restitution $V_{after}/|V_{before}| = \beta = 0.5$ (further details are provided at start of next section). These values of $\mu$ and $\beta$ were used in the model calculations.

The measured effects of the impact-induced tension as a function of the surface inclination angle are shown in Fig. 3. The top plot shows $\Delta z$ = height difference between the two chains just before the last dowel strikes the surface, as measured between the middle of the top dowels. The bottom plot shows the speed of the top rod just before it impacts the surface. $\Delta z$ was calculated from the final video frame before impact while the speed, $V$, was calculated by fitting a line to the positions obtained in the last 4 frames of the video before impact. For comparison, the free-fall speed at the surface height was $V_{freefall}$ = 5.08 m/s, which is approximately the lower cut-off of the vertical axis of this speed plot.

The number of measurements made at each inclination angle ranged from 11 to 17. To minimize systematic errors, the uniform-density chains were alternated between each run. The error bars on the measurements of $\Delta z$ extend away from the average value by one-half standard deviation above and below. These statistical variations are relatively small at all inclination angles, often smaller than the symbol size. In contrast, the error bars on the measurement of the speed are a combination of the statistical and estimated systematic uncertainties. The presence of the impact-induced tension is apparent at all measured inclination angles, 0 to 80 degree, since $\Delta z > 0$ and $V > V_{freefall}$.

The motion during a single fall is shown in Fig. 4 for the chain whose density decreases with height. This plot tracks the motion of the top rod of the two chains from before the release time (upper-left on the plot) through to a time two frames before the top rod is observed to strike the surface (bottom-right on the plot). The configuration of this experiment was designed to maximize the impact-induced tension effects: the variable-mass chain was used and the surface was inclined at an angle of 40°, which was the angle where Fig. 3 showed the maximum effect. The speed was calculated at each position by fitting a line to the position data starting two frames before to two frames after the position. No uncertainties were included in these position and speed values. This run corresponded to a final separation distance of $\Delta z$ = 17 cm, which is significantly larger than the maximum $\Delta z$ reported in Fig. 3 because here the variable mass chain was used. The large increase in the speed of the top rod just before it strikes the surface is apparent.

## III. THEORETICAL MODEL OF CHAIN-SURFACE INTERACTION

To theoretically model these experiments, a rigid body model is used to describe the collision between each rod of the chain and the surface. The collision is assumed to be instantaneous and effective parameters are used to describe the impulses that occur during the collision. A coefficient of friction and coefficient of restitution are used to describe the impulses parallel and perpendicular to the surface, respectively. This is a very popular approach; it has been used in models of granular systems, robotics, biomechanics and many others.[34,35] However there are multiple ways to define a coefficient of restitution and various unresolved issues associated with them.[34,36] Here, the coefficient of restitution is defined to be the ratio of the post-impact normal velocity to the pre-impact normal speed at the end of the rod that collides with the surface---which is usually called Newton's method. This method is relatively simple, is easy to understand, and it does a good job of describing the collision between a freely falling uniform metal rod striking a horizontal metal surface at shallow angles of tilt.[36] Here this model is used to calculate the impact-induced tension for a variety of conditions. While the model is not guaranteed to always be accurate, it is useful for understanding the general behavior of how the impact-induced tension is created, for explaining the results of the experiments, and for predicting the results of future experiments.

**A. Long chain striking a horizontal surface**

When one rod of the chains shown in Fig. 2 strikes a horizontal surface, there are two forces acting upwards on it; the force at the contact point between the rod and the surface, $N$, and the force at the other end of the rod from the string attached there to the chain above, $F$. By Newton's third law, $F$ is also the magnitude of the force acting on the chain above the rod, pulling it downward---the impact-induced tension. Fig. 5 shows the force diagram for the more general case of a rod striking an inclined surface; for the case discussed here $\theta = 0$ so $N$ is parallel to $F$ and the friction force $f$ is negligible. An expression for $F$ can be derived by applying Newton's laws of motion to the bottommost rod.

$$m\frac{dV_z}{dt} = F + N \quad (3)$$

$$I\frac{d\omega}{dt} = [F - N]\frac{L}{2} \quad (4)$$

Eq. (3) describes the center-of-mass motion of the rod; with $V_z$ the velocity of the center of mass in the vertical direction ($z$-direction in Fig. 5), $m$ the mass of the rod, and upwards is taken as the positive direction. Eq. (4) describes the rotation of the rod around a horizontal axis going through the center of mass ($y$-direction in Fig. 5); $\omega$ is the corresponding angular velocity of the rod; $L$ is the length of the rod; $I$ is the rotational inertia; and counterclockwise is taken to be the positive rotation direction. Eq. (4) assumes that the center of mass of the rod is at the middle of the rod; that the angle between the rod and the surface is small; and that the forces $F$ and $N$ act at the ends of the rod. The rod is assumed to be stiff and the surface hard, so that the change in lever arm length during the collision is negligible.

Rigid body parameters are used to describe the change in motion of a rod during the collision. In particular, the right and left ends of the rod are assumed to have the following velocities before and after the collision.

Left end, z dir.:   $-V \rightarrow +\beta V$     (5)
Right end, z dir.:  $-V \rightarrow -V$           (6)

Initially the rod is assumed to be falling with speed $V$ and no rotation. Assuming the velocity of the right-end to be unchanged during the collision is equivalent to assuming that the chain above has infinite mass and infinite rotational inertia---so that its change in motion during the collision is negligible---and also that the string connecting the rod to the chain above does not significantly stretch during the collision. Direct observation of the collision of the bottommost rod with the surface show that this parameterization is reasonable with a value of $\beta = 0.5$ for the materials used in our experiments.

From Eqs. (5) and (6) it is straightforward to derive that during the collision the change in the center of mass velocity is $\Delta V_z = (1+\beta)V/2$ and the change in the rotational velocity about the center of mass is $\Delta\omega = -(1+\beta)(V/L)$. Using these values, and Eqs. (3) and (4), the impact force on the rod from the surface, $N$, can be eliminated from the calculation, giving an expression for $J = \int F dt$ = the impulse on the chain above from the collision of the rod with the surface

$$J = \alpha m V \tag{7}$$

where $m$ is the mass of the bottommost rod and

$$\alpha = \left(\frac{1+\beta}{4}\right)[1-\xi] \tag{8}$$

Here $\xi = 4I/mL^2$ is a dimensionless measure of the rotational inertia of the rod, with $I$ the rotational inertia of the rod about the center of mass and $L$ the length of the rod. The impulse from the bottommost rod, Eq. (7), can be turned into an expression for the average force by dividing this impulse by the time between collisions of successive rods with the surface $= s/V$, where $s$ = the distance between the rods. This gives Eq. (2) for the average force, $F = \alpha\lambda V^2$, since $m/s = \lambda$ = the linear mass density. The form in Eq. (2) was justified in the introduction using dimensional analysis, but here it has been derived explicitly from Newton's laws. This derivation has allowed us to determine the value of $\alpha$ in terms of parameters describing the structure of the rod and the nature of the collision between the rod and the surface.

The general form for $\alpha$ in Eq. (8) can be understood from simple considerations. First note that $\alpha$ must vanish when no collision occurs, and this corresponds to when $\beta = -1$. Also, $\alpha$ must vanish when $\xi = 1$ since then the center of percussion of a rod about the right end (the end attached to the chain above) occurs at the left end of the rod where the collision with the surface occurs. For constant $\beta$, $\alpha$ is largest when $\xi = 0$, which is to be expected since a vanishing moment of inertia would tend to give rise to the largest possible rotations of a free rod and so here corresponds to the largest possible impact-induced tension. However, as is shown in the next section, the behavior in the $\xi \to 0$ limit is sensitive to the assumptions about what is above the rod colliding with the surface.

For a rod that is a uniform (which is the case for the chains used here) then $\xi = 1/3$. If the rod end remained on the surface after the collision then $\beta = 0$ and Eq. (8) gives $\alpha = 1/6$, which agrees with the analysis in Ref. 12. Using the value of $\beta$ that was directly observed in the experiments here, $\beta = 0.5$, then Eq. (8) gives $\alpha = 0.25$. This value is smaller than the energy conserving value of $\alpha = 0.5$, so it lies in the physically allowed range. This value is large compared to the value measured for a bead chain falling onto a horizontal surface in Ref. 22, which is 0.08 (when averaged over the bottom and middle parts of the bead chain). Thus it is to be expected that motion enhancement effects are relatively large for the type of chain studied here.

Unfortunately, the situation considered here is experimentally difficult to realize. This is because when the surface is horizontal the rods will collect into a pile. This pile tends to be rather irregular and yielding, and so the assumption of rods colliding with a flat, hard, horizontal surface is only valid for the first rod to collide with the surface. Before looking for a way around this problem, we shall first consider another important effect.

### B. Short chain striking a horizontal surface

In order to model what happens when the top end of the chain approaches the surface, the previous calculation must be extended to account for only a finite amount of chain above the rod colliding with the surface. This is done by now allowing the velocity of the right end (see Fig. 5, $\theta = 0$) of the bottommost rod to change during its collision with the surface. The size of this velocity change depends on how the chain above the bottommost rod responds to the pull downwards on the right-hand side.

All of the rods are attached to each other by strings at their ends, so a pull on one end causes all of them rotate together with the same angular velocity about their centers of mass. Thus the motion of the rods above the bottommost link can be modeled as a single effective rod with total mass $M$ and rotational inertia $I'$ about the center of mass. It is assumed that initially all of the rods have zero angular velocity, which is reasonable since the angular velocity induced in the rods above the bottommost by one collision will be approximately canceled by the next collision. Also, it is assumed that the end of the bottommost rod that collides with the surface rebounds with the impact parameter $\beta$. Then applying Newton's laws to both the effective rod and the bottommost rod[37] gives the impact-induced impulse

$$J = \frac{\alpha m V}{1 + \gamma [m/M]} \tag{9}$$

where $\alpha$ is the same as Eq. (8) and

$$\gamma = \frac{1}{4}(1 + \xi)\left(1 + \frac{1}{\xi'}\right) \tag{10}$$

Here $\xi' = 4I'/ML^2$ is the dimensionless rotational inertia associated with the part of the chain above the bottommost rod. If the chain's rods all have similar mass distributions about their centers of mass, then $I'=(M/m)I$ and so $\xi' = \xi$. This is the case for the chains experimentally studied here; however the distinction in the $\xi$'s is kept in Eq. (10) to help identify how the different rotations influence the motion.

The impulse from the bottommost rod given in Eq. (9) can be turned into an expression for the average force by dividing the impulse by the time $s/V$ between collisions of each rod with the surface. Using that $m/s = \lambda(z) = $ the density of the chain near the surface, a distance $z$ from the end, then the force can be written as

$$F(z,V) = \frac{\alpha\lambda(z)V^2}{1+\gamma[m(z)/M(z)]} \qquad (11)$$

Here $m(z)$ is the mass of the rod near the surface and $M(z) = \int_0^z \lambda(z)dz$ = the total mass of the chain above the bottommost rod. When the chain has uniform mass density, then $m/M = s/z$. Eq. (11) is a more general expression for the impact-induced tension than Eq. (2), since it holds for all values of $z$ while Eq. (2) is only correct when the end of the chain is far from the surface, $M \gg \gamma m$.

The general form for the impact-induced tension given in Eq. (11) can be readily understood. When one end of the bottommost rod strikes the surface, the other end pulls downward on the chain above, creating the impact-induced tension. The more the chain above moves in response to that pull, the smaller the impact-induced tension. This is apparent in the results above since $F$ as given by Eq. (11) is always less than or equal to $F$ as derived in section III.A and given in Eq. (2). Equality holds when the chain above is very long (for nonzero $\xi'$) since then $\gamma(m/M) \ll 1$ and we recover the result derived in section III.A. In the opposite limit, the impact-induced tension vanishes when the chain above moves easily and there is nothing for the bottommost rod to pull on. This happens when $\gamma m/M \gg 1$, which gives $F \to 0$. There are two different physical situations when this limit can happen. The first situation is when the chain end approaches the surface, so $M \to 0$. When there is no mass to pull against, the tension vanishes. This is why the tension at the top end of the chain is always zero. The other situation is when the rotational inertia about the center of mass of the rods above the bottommost rod vanishes, $\xi' \to 0$. The force $F$ is applied at the end of the rods above the bottommost, so if the rods above have no rotational inertia they will rotate freely and thus $F$ will vanish.

The vanishing of the impact-induced-tension as the end of the chain reaches the surface has important implications when trying to solve for the velocity of the chain at this point. In particular, from Eq. (1) it is easy to see that if $F$ were nonzero as the chain length, $z$, approaches 0, the acceleration would becomes infinite due to the vanishing of the mass. This also means the velocity of the chain at this point would be infinite. Instead, using Eq. (11) for $F$ in Eq. (1) leads to a finite velocity at the end of the chain.

The combination of the rotational inertias given in Eq. (10) will be seen again in the next section, in a different context. It is explained in more detail there.

### C. Long chain striking an inclined surface

Chain collisions with a horizontal surface are difficult to realize experimentally because the chain collects into an irregular pile. However if the surface is sufficiently inclined, rotated about a horizontal axis parallel to the width of the chain, then a pile will not form because the chain will smoothly slide away from the collision point. Then single rod-surface collisions should accurately describe the creation of the impact-induced tension.

Here the calculation in section III.A is extended to the more general case of an inclined surface, $\theta \neq 0$ in Fig. 5. Now the change in the rod's speed parallel to the surface is relevant and depends on the friction force, $f$. This is taken to be $f=\mu N$, where $\mu$ is the coefficient of friction. This is an approximate treatment of friction as it ignores the effects of static friction. This simplification is intended to capture the dominant contribution of friction at large inclinations, while yielding a relatively simple analytical expression for the impact-induced tension.

Newton's laws of motion are applied to the collision of the bottommost rod with the surface.[37] It is assumed that the rods are symmetric such that the rotational inertia matrix is diagonal. This is reasonable for the experiments performed here since the colliding object is a uniform rod, so that the rotational inertia about the center of mass is the same about the $y$- and $z$- axes. However the subscripts on the rotational inertia are kept in what follows to help identify the contribution of the different dynamics to the final expression for the impact-induced tension. The diameter of the rod is assumed to be small so that rotation of the rod about the $x$-axis is negligible. Also, the angle between the surface and the long axis of the rod is assumed to be small so that motion along the $x$-axis is negligible. It is also assumed that the center of mass of the rod is at the middle of the rod, and that the forces $f$, $F$ and $N$ act at the ends of the rod.

The equations of motion can be solved to find the impulse from the impact-induced tension, $J$. The form for the impulse is the same as given in Eq. (7), but the expression for $\alpha$ is different.

$$\alpha = \left(\frac{1+\beta}{4}\right)[1-\xi]\frac{\cos^2\theta}{g(\theta)} \tag{12}$$

where $g(\theta)$ is given by

$$g(\theta) = \left[\cos^2\theta + \sin^2\theta\frac{1}{4}\left(1+\frac{1}{\xi_z}\right)(1+\xi_y)\left[\frac{1-\mu\cot\theta}{1+\mu\tan\theta}\right]\right]. \tag{13}$$

Here $\xi_y$ and $\xi_z$ are the dimensionless rotational inertias ($4I/mL^2$) about the $y$- and $z$-axes going through the center of mass.

The general form of Eqs. (12) and (13) can be readily understood. When $\theta \to 0$, these equations reduce to the expression in Eq. (8) for a horizontal surface. Eq. (12) is the same as Eq. (8) except for the additional factor of $\cos^2\theta/g(\theta)$. The common factors are there for the same reasons given in section III.A. In addition, for vanishing friction ($\mu=0$), and assuming a uniform rod then $\xi = 1/3$, so $g(\theta) = 1+(1/3)\sin^2\theta$, which is a slowly varying function of $\theta$. Thus the dominant effect of the surface inclination is the factor of $\cos^2\theta$. This factor is what one would tend to guess beforehand, since the component of the normal force in the $z$-direction is smaller by a factor of $\cos\theta$, and also the change in velocity in the $z$-direction is smaller by the same factor.

It is interesting to consider more closely how $\alpha$ depends on the dimensionless rotational inertia, $\xi$. In particular, as the rotational inertia about the $z$-axis goes to zero ($\xi_z \to 0$) then $g(\theta) \to$ infinity and so $\alpha \to 0$ and the impact-induced tension vanishes. This is very similar to the behavior found in section III.B, where it was noted that the impact-induced tension vanishes when $\xi' \to 0$. The two effects are closely related, as can be seen by noting that the combination of rotational inertias as defined as $\gamma$ in Eq. (10) is structurally identical to the combination in Eq. (13). This common dependence on the $\xi$'s occurs because for both cases a force is applied to the end of a *free* rod, producing rotational motion and translation. In section III.B, the rods freely rotating are the rods above the bottommost, while for the calculation in this section the freely rotating rod is the bottommost rod rotating about the axis in the z-direction. The vertical string attached to the bottommost rod does not constrain the motion of this end of the rod in the horizontal (y-direction) during the brief collision, thus for rotation about the z-axis during the collision the rod is essentially free. The smaller that $\xi_z$ is, the larger the rotation produced about the z-axis and also the larger the center of mass speed produced in the y-direction by the collision. There is thus less energy left to be recycled into the vertical direction and thus the impact-induced tension is decreased.

It is also interesting to examine closely the effects of friction. Eqs. (12) and (13) show that friction always makes the impact-induced tension larger, with the largest increases being at steep inclinations. This is because friction increases the size of the upward component of the force acting at the end of the rod colliding with an inclined surface (see Fig. 5), and it is this upward force that creates the impact-induced tension. At very steep inclinations the upward component of the friction force can be larger than the upward component of the normal force, even for small coefficients of friction. In particular, at $\theta$ = 80° (the steepest inclination observed here), and $\mu$ = 0.2, friction effects increase the size of $\alpha$ by a factor of 2.25. At $\theta$ = 40° the increase in $\alpha$ is only about 20%. The corresponding increases in the separation distance and the speed at the end of the chain are approximately proportional to the increase in $\alpha$, as discussed in the next section.

**D. Predictions for chain lay-down**

The previous calculation can be extended to account for only a finite amount of chain above the surface---as was done in section III.B for a horizontal surface. However for an inclined surface there are now three different types of disturbances acting on the end of the chain stretching upwards that must be considered. One is the impulse in the vertical direction, and the effect of this has already been calculated in section III.B. The other two disturbances are the creation of a transverse pulse (motion along the y-axis) and a torsional pulse (rotation about the z-axis). These latter two disturbances are excitations of new degrees of freedom that are present because of the inclined surface.

The effects of the transverse and torsional disturbances on the chain above are expected to be small. This is because the speed of the chain downward, $V$, is faster than the wave speed of these disturbances. Thus they cannot propagate upwards to alter the chain's motion, and the assumption of an infinite chain, as made in Section III.C, should be reasonable. For example, the speed of transverse pulses on the chain (motion along the y-axis) is approximately $V_{\text{transverse wave}} = \sqrt{F/\lambda} = \sqrt{\alpha}V \approx \sqrt{0.1}V < V$ where Eq. (2) has been used for the tension[11,26]. Similarly, the speed of torsional waves on the chain (twisting around the z-axis) is $V_{\text{torsional wave}} = \sqrt{F/\xi\lambda} = \sqrt{\alpha/\xi}V \approx \sqrt{0.3}V < V$. In contrast, the impulses applied to the chain in the vertical direction are qualitatively different because the chain responds instantaneously to these forces (for a stiff string) and thus they must be accounted for. Therefore the form of the impulse found in section III.B should also be valid for an inclined surface.

The solid curves in Figs. 3 and 4, and both curves in Fig. 6, show the predicted behavior for the speed and positions of the falling chains. These predictions were calculated using Eqs. (9) and (10) from section III.B with Eqs. (12) and (13) from section III.C to give the increase in speed of the chain that occurred when each rod collided with the surface, $\Delta V = J/M$. In between these collisions the standard free-fall equations were used to describe the motion. Small effects, such as the tilt of the rods from horizontal and the vertical width of the rods, were neglected. These calculations were implemented using a spreadsheet. The coefficients of restitution and friction were taken to have the values found in direct measurements of these quantities, $\beta = 0.5$ and $\mu = 0.2$ (see section II), thus there are no free parameters in these predictions.

For Fig. 3, the agreement with the experimental measurements is reasonably good for inclinations angles greater than 30° especially considering that there are no free parameters in the fit. For 30° and smaller the rods in the chain collided primarily with a pile of rods on the surface and not directly with the surface. For that region the surface was assumed to be horizontal ($\theta = 0$) and an effective coefficient of restitution of $\beta = -0.37$ was chosen to best fit the data, resulting in the dashed, horizontal lines on Fig. 3. This negative value of $\beta$ is consistent with the idea that the pile of rods is an irregular and also yielding surface. The angle at which the transition occurs from collisions on a pile to collisions on the surface is not easily calculable. This is because a single rod will collide multiple times with the surface before it starts moving down the surface, and the size of the forces acting on the rod during this process is sensitive to small effects.

Fig. 4 shows the height of the chain as a function of speed-squared, for the chain whose density decreased with height, at an inclination of $\theta = 40°$. The agreement between the experimental measurements and the theoretical predication is excellent. This is not too surprising since the agreement between theory and experiments was also very good in Fig. 3 at the inclination angle of 40° for a chain with uniform density. It is clear from Fig. 4 that the effects of the impact-induced tension are largest when the end of the chain nears the surface. This is consistent with theoretical expectations because the force acting on the chain is proportional to the speed-squared (see Eq. (2)) and because the acceleration of the chain is approximately inversely proportional to the mass of the chain above the surface (see Eq. (1)).

Fig. (6) shows the predicted dependence of the separation distance, $\Delta z$, and the speed at the end of the chain as a function of the rotational inertia about the center of the rod. Here it was assumed that all of the rods in the chain were identical ($\xi = \xi'$ in Eq. (10)) and symmetric around the x-axis ($\xi_y = \xi_z$ in Eq. (13)). The inclination angle was taken to be $\theta = 40°$, and the coefficients of restitution and friction found in this experiment were used ($\beta = 0.5$ and $\mu = 0.2$). The plot shows that the impact-induced tension effects vanish in the limits of $\xi \to 1$ (as discussed in section III.A) and also when $\xi \to 0$ (as discussed in sections III.B and III.C). The general dependence of the separation difference, $\Delta z$, and the speed on $\xi$ are approximately the same, but not exactly since the two curves peak at slight different values of $\xi$. For the experimental results given in Section II, the rods were uniform so $\xi = 1/3$. This value is close to the peaks in both $\Delta z$ and speed, and so only a small increase in impact-induced tension effects would be possible from changing the rotational inertia of the rods. Given these predictions such experiments were not attempted here.

The numerical simulation predicts the motion of the falling chain by adding the impulses from when each link hits the floor to the free-fall motion between collisions. An alternative method of predicting the motion is to solve Eq. (1). This method provides a check on the numerical simulation, a check on Eq. (1), and also gives insight into the motion.

It was shown in Ref. (26) that Eq. (1) with the form of the force in Eq. (2) can be solved analytically for a constant density chain. The methods used there can also be used with the more general form of the force given in Eq. (11).[37] These methods yield an equation for the speed-squared of the chain as a function of the height, $z$.

$$V^2(z) = \frac{2g}{[z+\gamma s]^{2\alpha}} \left\{ h[z_0 + \gamma s]^{2\alpha} + \frac{1}{1+2\alpha} ([z_0 + \gamma s]^{1+2\alpha} - [z + \gamma s]^{1+2\alpha}) \right\}. \tag{14}$$

Here $z_0$ is the initial length of the chain and $h$ is the distance the chain falls before it first strikes the surface. Eq. (14) is an exact solution for the speed of a constant density chain as a function of height. The speed when the top end of the chain strikes the surface is found by evaluating Eq. (14) at $z=0$. The comparison of this result to actual measurements is slightly ambiguous as this is a continuous approximation to what is a discrete process, and thus it is unclear whether it applies before or after the last rod collides with the surface. Applying it to before this collision, Eq. (14) reproduces the numerical simulation results shown in Fig. 6 to within 4%. Also, it is interesting to note that taking the separation distance between the rods to vanish ($s \to 0$) results in the speed at the end growing logarithmically to infinity, which agrees with the discussion in section III.B.

For the experiments discussed here, $\alpha$ is of order 0.1 or less and $h/z_0$ is 0.15. Thus it is numerically reasonable to take $h = 0$ and also to work to just leading order in $\alpha$. Then the separation distance at the end can be calculated[37] using the methods given in Ref. (26) to be

$$\frac{\Delta z}{z_0} \approx 2\alpha \left\{ \sqrt{1 + \frac{\gamma s}{z_0}} \ln\left[ \left(\sqrt{1 + \frac{\gamma s}{z_0}} + 1\right)^2 \left(\frac{z_0}{\gamma s}\right) \right] - \left(1 + \frac{\gamma s}{z_0}\right) \ln\left[1 + \frac{z_0}{\gamma s}\right] - 1 \right\}. \tag{15}$$

This equation successfully reproduces the numerical calculation for $\Delta z$ shown in Fig. 6 to within 4%. In the limit of $\gamma s/z_0 \to 0$, this equation reduces to $\Delta z/z_0 \approx 2 (2\ln 2-1)\alpha$, which is the same result as found in Ref. 26. In the limit $\gamma s/z_0 \to \infty$ then Eq. (15) gives $\Delta z/z_0 \to 0$ which agrees with expectations as explained in section III.B.

It is interesting that the expression for $\Delta z$ in Eq. (15) is independent of $g$, the acceleration of gravity. This is not an artifact of the approximations made to derive this solution, or the result of assuming constant density, but is in fact exactly true for all mass distributions. This result can be proved using a dimensional scaling argument with Eqs.

(1) and (11). Hence if the experiments were repeated on the moon, the same values of $\Delta z$ would be measured there as on the Earth.

## IV. DISCUSSION

There have been many studies of motion enhancement in falling chains. These works generally avoid theoretical discussions of what happens when the mass of the falling chain becomes small because then the structure of the chain is important. However, understanding this situation is necessary for calculating the speed at the end of the falling chain and for modeling the chain with decreasing density. For the chain type studied here, one made of slightly tilted rods, calculating the impact-induced tension is relatively straightforward, and the behavior when the mass of the falling chain is small is described by Eq. (11). While the parameters $\alpha$ and $\gamma$ in this equation are specific to this chain type, it is likely that the general form of this equation applies to many other situations, such as the fall of a bead chain onto a surface, and also to the fall of a folded chain.

A few experimental groups have observed motion enhancement effects in bead chains falling onto a surface.[11,22] The structure of bead chains is very different than the structure of the chains used here, and, not surprisingly, the motion enhancement effects observed in these chains are also qualitatively different. For example, for bead chains the separation distance at the end of the chain, $\Delta z$, is observed to vanish at intermediate surface inclination angles around $\theta = 52.5°$.[22] In contrast, for the experimental results presented here, there is no vanishing of $\Delta z$ at intermediate angles, and in fact the value at $\theta = 50°$ is one of the largest (see Fig. 3). This demonstrates that motion enhancement can depend sensitively on the internal structure of the chain.

The explanations of the motion enhancement are qualitatively different for the two types of chains. For bead chains the impact-induced tension is proposed to be caused by the transverse, pendular motion of a bead about the bottom of the chain,[26] and therefore it occurs over a finite time. It is typically rather fragile and difficult to model because it is sensitive to how the collision with the surface creates the transverse motion. In contrast, in the chain discussed in this paper, the impact of a rod with the surface directly creates the tension and is therefore relatively robust.

For an arbitrary type of chain or string, it is possible for both mechanisms to contribute to producing the impact-induced tension. For example, let's consider chains of the type discussed here, but with a large angle between a rod and the horizontal. Then, after one end of the rod collides with the surface, the rod will rotate upwards about the other end. During this rotation there is a downward force on the string connecting the rod to the rest of the chain. The impulse from this rotation is small when the tilt angle of the rods is small---as is the case for the chains considered here. However this small contribution to the tension from the pendular mechanism may explain why many of the predicted values in Fig. 3 are slightly below the observed values.

While the experiments reported here used a somewhat high-speed video camera and video analysis to track the relative motion of the chains, there are easier and less expensive ways of doing the experiments. For example, a photogate or other type of photo switch could be used to observe a falling chain of the type studied here, one made of slightly tilted rods. Then the experiment is quite similar to the common introductory physics lab activity of observing the motion of a picket fence falling through a photogate. While the falling picket fence is used to demonstrate free-fall, a falling chain of rods demonstrates the physics of a variable mass system. Such experiments could readily be performed as an introductory physics lab activity, or as part of an undergraduate research project.

There are a plethora of possible research projects that could be done on falling chains striking a surface. For chains made of rods connected by string at each end (like those studied here) some unanswered questions are, "How does motion enhancement depend on the tilt angle of the rods?" and "Can the chain fountain be observed for this type of chain?" and "How accurate are the predictions in Fig. 6 for how the motion depends on the rotational inertia of the rods?" For chains of any type whose mass density decrease with height, some interesting questions are "What type of mass distribution optimizes the motion enhancement effects?" or "How long of a chain does it take to break the speed of sound?" More general questions of interest are, "Are their other mechanisms for producing the impact-induced tension besides the "impact" mechanism and the "pendular" mechanism?" and "How large is the motion enhancement in strings?"

The chain type studied here, made of rods that are slightly tilted from horizontal, is rather artificial---it is not a part of standard technology. However this chain is useful for demonstrating the existence of the impact-induced tension and for understanding how this tension can be produced. For the strings and chains used in technological applications, it seems likely that many, if not most, will create a motion enhancing force when they interact with a surface. The size of this force depends on the relative speed and on the internal structure of the chain or string. Understanding this force could lead to more efficient technology, and possibly also to new applications for strings and chains.

## ACKNOWLEDGEMENTS

The author would like to thank his wife Sharyl and his son Rob for their patience while the family home was used to perform all of the experiments presented here---school buildings were closed due to the pandemic. Thanks also go to Charles Panigeo for his interest in falling chains, and for his preliminary studies of falling chains whose density decrease with height.

*jtpantaleone@alaska.edu

**FIGURES**

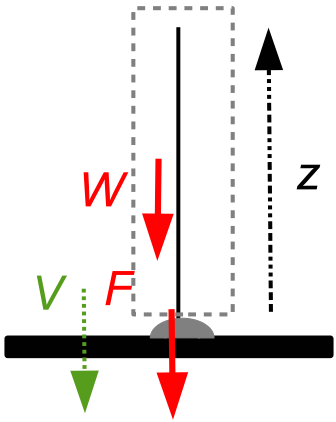

Fig. 1. Two forces act on the section of the falling chain inside the dashed region: the weight, $W$, and the impact-induced tension, $F$, acting at the bottom of the dashed region. $V$ is the velocity and $z$ is the height of the chain.

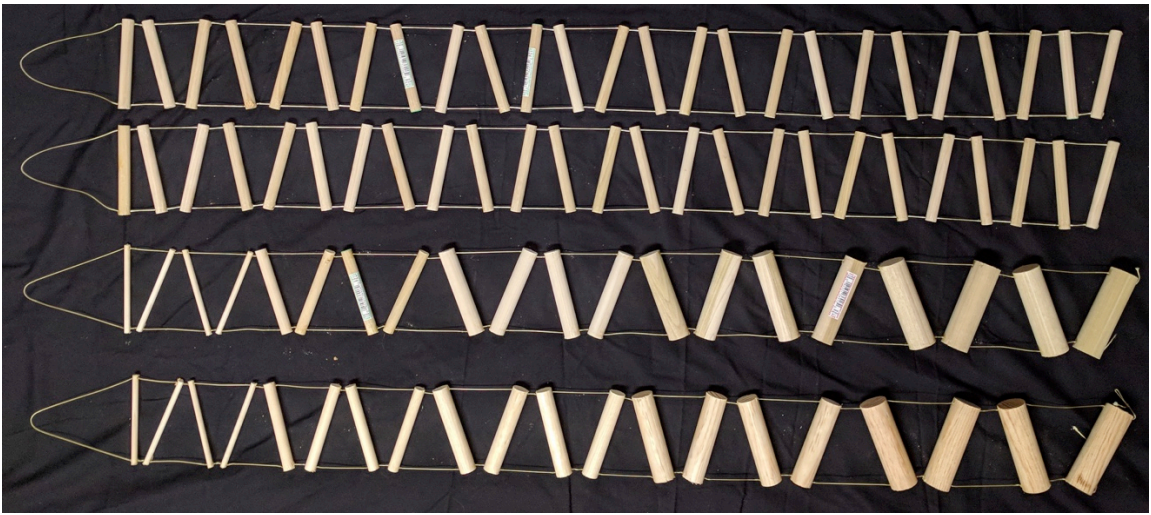

Fig. 2. Photograph of the chains used in these experiments. The bottom two chains were used to study motion enhancement in a variable mass density chain. The top two chains were used to study the effects of an inclined surface.

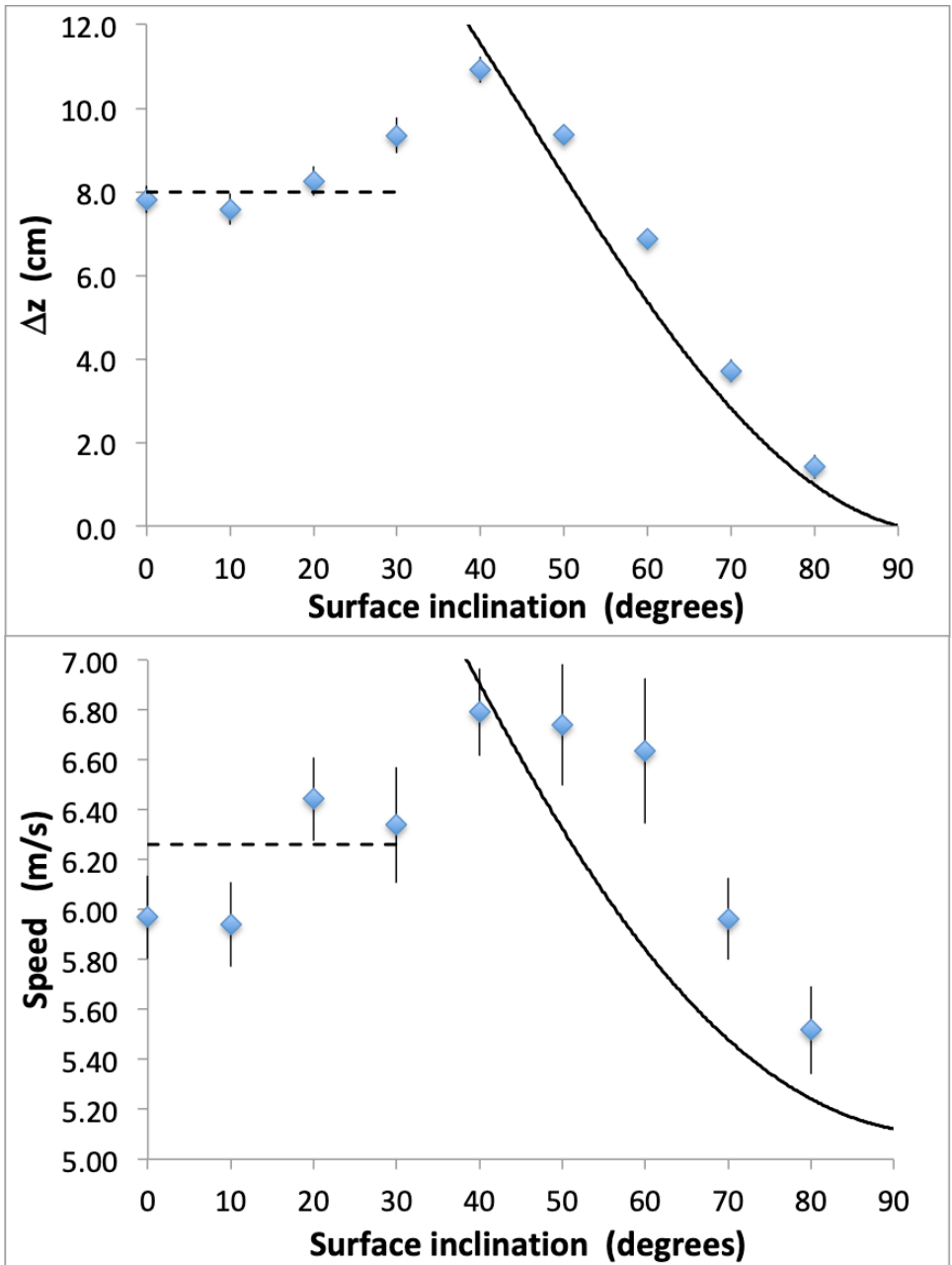

Fig. 3. Separation distance between the two chains (top) and speed of the top rod (bottom) just before it hits the surface, as a function of surface inclination with respect to the horizontal. Markers are measured values and curves are model predictions. The solid curve has no free parameters, while the dashed curve has one fit parameter to model the "softness" of the pile of rods.

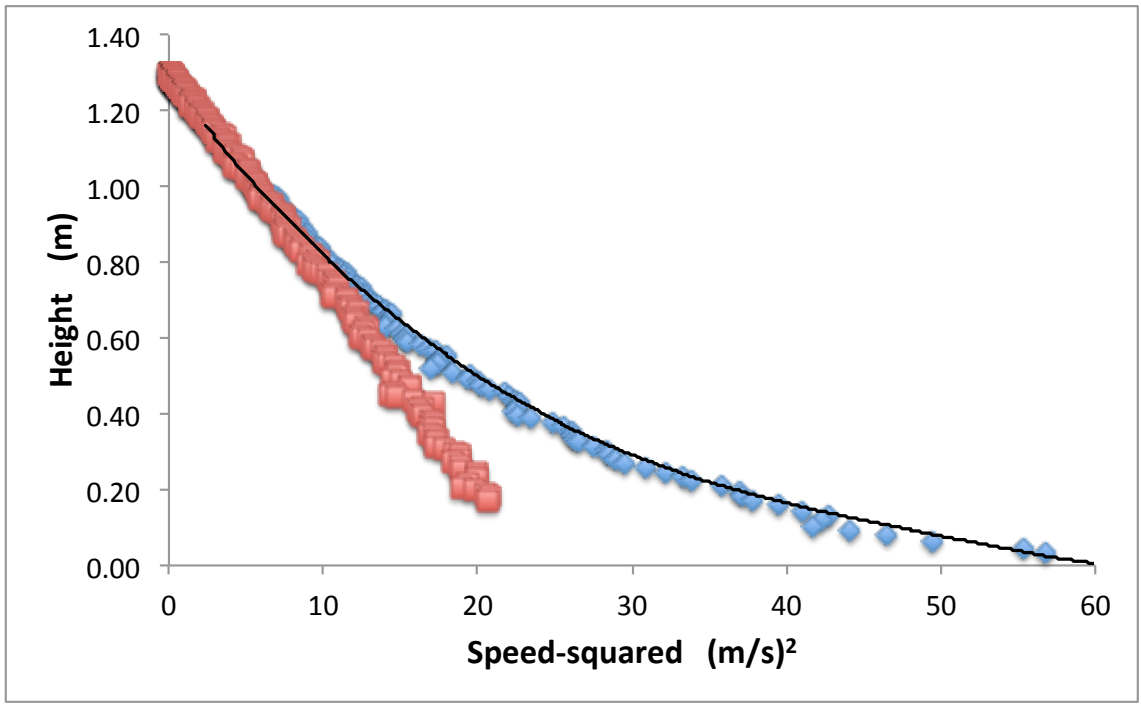

Fig. 4. Plot of height versus speed-squared for chains whose density decreases with height. Pink square markers are for the chain in free-fall; blue diamond markers are for the chain striking the surface; and the solid curve is the model predictions (with no free parameters). The surface was inclined at 40°, and the final separation distance just before impact was $\Delta z = 17$ cm.

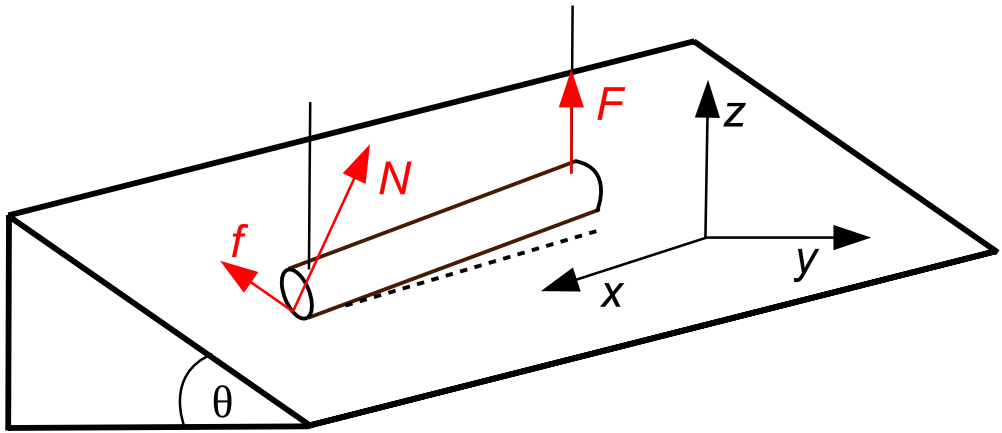

Fig. 5. Diagram of the forces acting on the bottommost rod when it strikes an inclined surface.

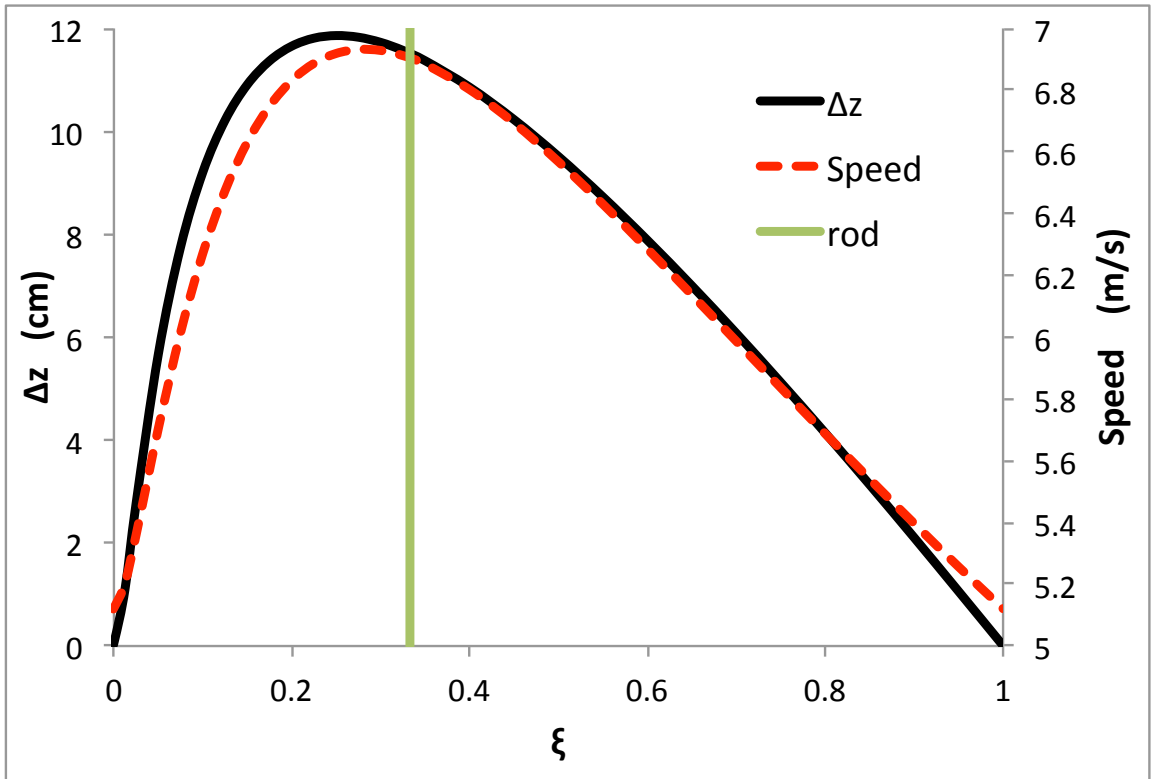

Fig. 6. Predicted effects of the impact-induced tension as a function of the dimensionless rotational inertia of the rods, $\xi$. The curves show the separation in chain heights, $\Delta z$, and the speed, just before the last rod collides with the surface. The surface is assumed to be inclined at $\theta = 40°$, $\beta = 0.5$, $\mu = 0.2$, and the chain density is uniform. The vertical line denotes the current experiments, with the rods having a uniform mass distribution so $\xi = 1/3$.

Supplementary material for "**Understanding and enhancing the impact-induced tension of a falling chain**" by J. Pantaleone.

This supplement contains additional details on the experiment, additional details on the many theoretical calculations, and a discussion of link lift-off from a horizontal surface.

**S I: Additional experimental details**

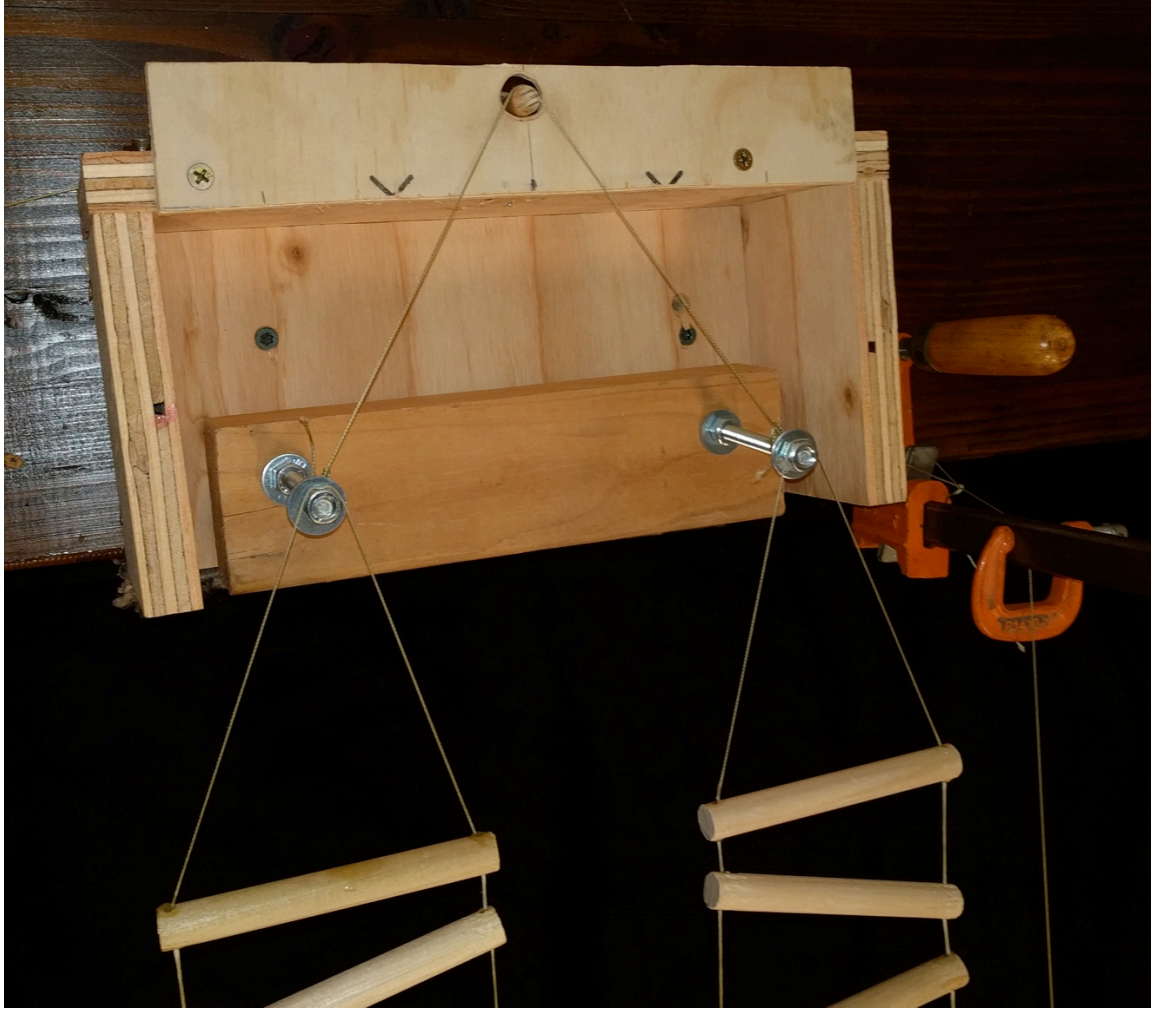

Fig. S1. Photograph of the mechanism used to release the falling chains. When the wooden peg at the top is retracted through the hole, the two metal rods rotate downward, smoothly releasing the chains.

The string used to construct the chains was 130 lb (58.9 kg) test Cortland "MasterBraid" fishing line with a diameter of 0.026 in (0.66 mm).

Several types of release mechanism were tried before settling on the design shown in Fig. S1. One type of mechanism tested was two, identical push-pull solenoids (the kind used in vending machines) wired in parallel. This mechanism was found to give rise to random differences in the relative release time of a few milliseconds. Such timing differences lead to fluctuations in the relative height of the freely falling masses of a few centimeters at a distance of 2 meters. This was judged to be excessive, so this mechanism was not used.

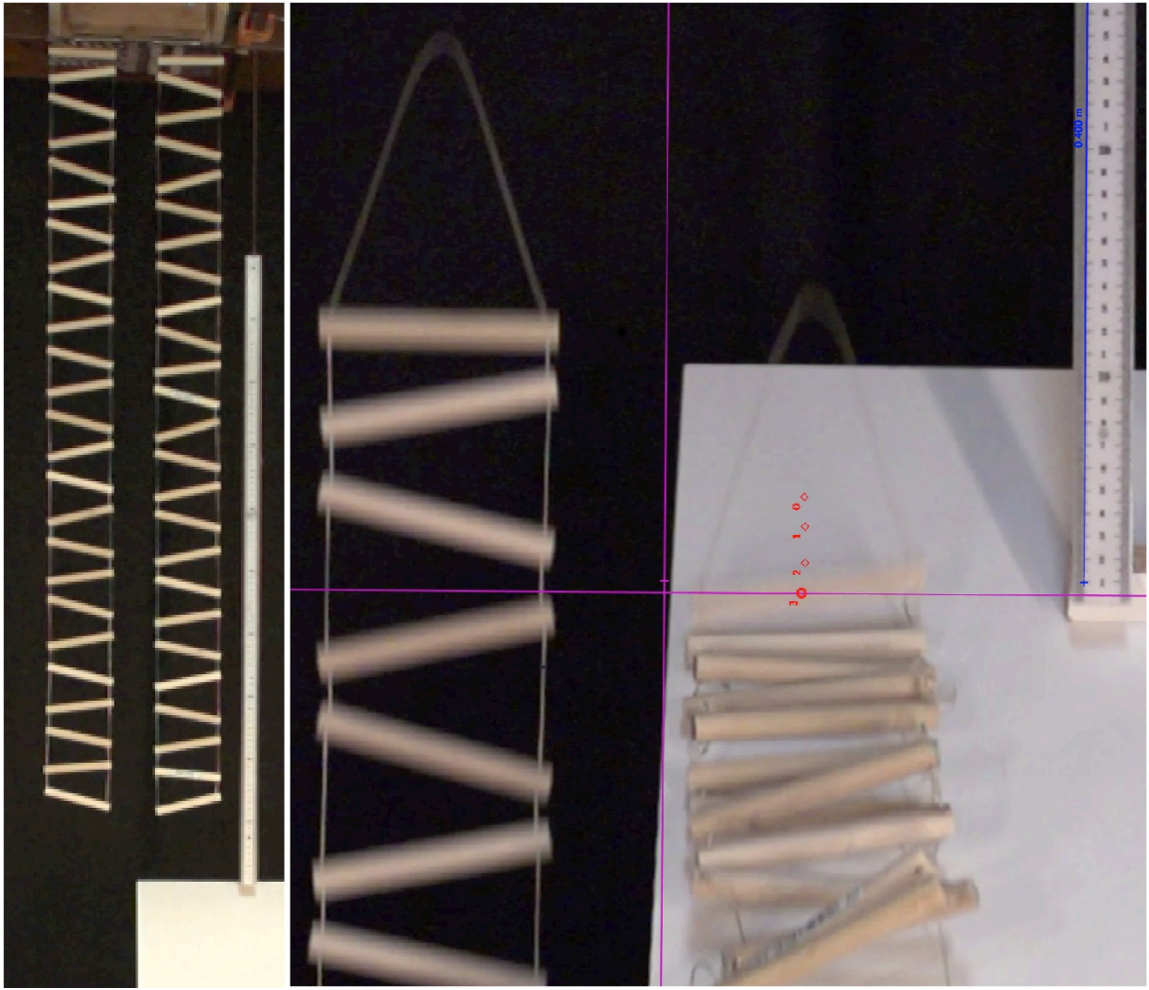

Fig. S2. Photographs showing the two chains before release (left) and a close-up of the chains just before the last rod strikes the surface (right). The surface inclination is 40° with respect to horizontal and the chains are uniform density. Part of a meterstick is shown at right and the location of the last rod in the last 4 video frames is marked with red diamonds. The last rod is moving rather quickly just before the collision, so its image is blurry.

**S II: Short chain striking a horizontal surface, details on the derivation.**

The calculation in Section III.A of the impact-induced tension for a long chain can be extended to account for a finite amount of chain above the rod colliding with the surface. The equations of motion describing the collision between the bottommost rod with the surface, Eqs. (3) and (4), remain the same as in section III.A, but now we include how the upper part of the chain responds to this collision. Applying Newton's laws of motion to the upper part of the chain gives

$$M \frac{dV_z'}{dt} = -F \quad \text{(S1)}$$

$$I' \frac{d\omega'}{dt} = -F \frac{L}{2} \quad \text{(S2)}$$

Eq. (S1) describes the change in center of mass motion of the upper part of the chain; with $V_z'$ the velocity of the center of mass in the vertical direction and $M$ the mass of the chain above the rod colliding with the surface. Eq. (S2) describes the effects of the torque from force $F$ on the rotation of the rods in the chain. Because all of these rods are attached to each other by string at their ends, all of these rods rotate together with the same angular velocity, $\omega'$. Thus the motion of these rods can be combined into that of a single effective rod with rotational inertia $I'$. Assuming all of the rods have the same geometric mass distribution, symmetric about the center, then $I'=(M/m)I$. Eq. (S2) assumes the tilt angle of these rods is small, and that the rods and string are stiff so that the changes due to the collision are instantaneous and also so that the change in lever arm length during the collision is negligible.

The changes in velocities of the rod-ends during the collision of the bottommost rod are constrained by the strings attached to them. They can be parameterized as

| | | |
|---|---|---|
| Bottom rod, Left end, z dir.: | $-V \rightarrow +\beta V$ | (S3) |
| Bottom rod, Right end, z dir.: | $-V \rightarrow -U_R$ | (S4) |
| Upper rods, Right end, z dir.: | $-V \rightarrow -U_R$ | (S5) |
| Upper rods, Left end, z dir.: | $-V \rightarrow -U_L$ | (S6) |

Initially, all of the rods are assumed to be falling with speed $V$ and no rotation. This assumption is reasonable since the rotation of the upper rods created by one collision are approximately canceled out by the next collision, so a cumulative rotation does not develop. The end of the rod that collides with the surface is assumed to rebound the same as before, i.e. Eqs. (S3) and (5) are the same. After the collision, the speed of the right end of the bottommost rod and the chain above will have the same value, $U_R$, because they are attached to each other by string, see Eqs. (S4) and (S5). The speed of the left end of the chain above after the collision is denoted as $U_L$. The change in center of mass velocities and rotations of the bottommost rod and the upper rods can be written in terms of the speeds given in Eqs. (S3) thru (S6).

Overall, there are four linear equations of motion (Eqs. (3), (4), (S1) and (S2)) and 4 unknowns: the impulses produced by forces $F$ and $N$, and the speeds $U_R$, and $U_L$. This system of equations can be solved to find the impact-induced impulse, $J$, in terms of known quantities. The result is given in section III.B by Eqs. (9) and (10).

**S III. Long chain striking an inclined surface, details on the derivation.**

The chain above the collision point is assumed to be long, i.e. that it does not change its motion due to the collision of one rod with the surface. Applying Newton's laws of motion to the collision of the bottom rod with an inclined surface, as shown in Fig. 5, gives for the generalized version of Eqs. (3) and (4)

$$m\frac{dV_z}{dt} = N\cos\theta + f\sin\theta + F \qquad (S7)$$

$$I_y\frac{d\omega}{dt} = \frac{L}{2}[-N\cos\theta - f\sin\theta + F] \qquad (S8)$$

Eq. (S7) describes the center of mass motion in the *z*-direction and Eq. (S8) describes the rotation about an axis going through the center of mass and parallel to the *y*-axis. In addition to these equations there is motion in new directions due to the inclination.

$$m\frac{dV_y}{dt} = N\sin\theta - f\cos\theta \qquad (S9)$$

$$I_z\frac{d\psi}{dt} = \frac{L}{2}[N\sin\theta - f\cos\theta] \qquad (S10)$$

Eq. (S9) describes the center of mass motion in the *y*-direction and Eq. (S10) describes the rotation about the axis parallel to the *z*-axis and going through the center of mass. Here, as in sections III.A and III.B, the coordinate system given in Fig. 5 has been used, *N* is the normal force and *F* is the tension in the string between the rod colliding with the surface and the rest of the chain. The friction force *f*, the component of the contact force parallel to the surface, is now relevant. The rotations of the rods are described by the angular velocities $\omega$ and $\psi$, and the rotational inertias $I_y$ and $I_z$, about the *y*- and *z*-axes, respectively. Eqs. (S8) and (S10) assume that the rods are symmetric such that the rotational inertia matrix is diagonal. This is reasonable for the experiments performed here since the colliding object is a uniform rod with $I_z = I_y$. However the subscripts on the rotational inertia are kept to help identify the contribution of the different dynamics to the final expression for the impact-induced tension. The diameter of the rod is assumed to be small so that rotation of the rod about the *x*- axis is negligible. Also, the angle between the surface and the long axis of the rod is assumed to be small so that motion along the *x*-axis is negligible. Eqs. (S8) and (S10) assume that the center of mass of the rod is at the middle of the rod, and that the forces *f*, *F* and *N* act at the ends of the rod. The rod is assumed to be stiff and the surface hard, so that the collision is instantaneous and thus the change in lever arm length during the collision is negligible.

Rigid body parameters are used to describe the change in motion of the bottommost rod during the collision. In particular, the following velocity component changes of the ends of the rod are assumed.

        Left end, perpendicular dir.:   $-V\cos\theta \to +\beta V\cos\theta$         (S11)
        Right end, *z* dir.:                $-V \to -V$         (S12)

Eq. (S11) is the generalized versions of Eq. (5), where the coefficient of restitution $\beta$ is used to describe the change in velocity of the rod end that collides with the surface in the direction perpendicular to the surface. Eq. (S12) is the same as Eq. (6) since the chain above is assumed to be infinitely massive and have infinite rotational inertia, so there is no change in speed in the *z* direction. Besides the velocity component values given in Eqs. (S11) and (S12), there are two other velocity components for the ends of the rod whose values after the collision are not known. In addition to these two unknown final velocity components, the other unknowns are the impulses from the forces *f*, *F* and *N*, for a total of 5 unknowns. With the four equations of motion, Eqs. (S7) thru (S10), one additional condition is needed to be able to solve for the impact-induced tension. This is taken to be $f = \mu N$, where $\mu$ is a coefficient of friction to be determined by experiment.

This is a simplified treatment of friction, as it ignores the effects of static friction. This approximate treatment of the friction force is intended to capture the dominant contributions of friction at large inclinations, while yielding a relatively simple, analytical expression for the impact-induced tension.

This system of equations can now be solved to find the impulse from the impact-induced tension, $J$, in terms of known quantities. Doing the algebra, one finds that the form for the impulse is the same as given in Eq. (7), but now the expression for $\alpha$ is different, and is given in Eqs. (12) and (13).

**S IV. Analytical solutions for the chain's motion.**

The theoretical predictions plotted in the figures are for a discrete chain. They were calculated by alternately adding free-fall motion to the impulses from each rod colliding with the surface. This numerical method can be used to describe the motion for any chain, but gives limited insight into the dynamics. Deeper understanding of the motion can be obtained by averaging over the chain's discreteness to get a differential equation for the motion, Eq. (1), and then solving this equation. Such an analytical solution exists for a constant density chain with the force as described by Eq. (2), see Ref. (26). Here this solution is generalized to the more realistic case where the impact-induced tension is described by Eq. (11). The latter gives a more accurate description of what happens when the end of the chain is near the surface, and is useful for understanding the general behavior of the falling chain as its parameters are varied.

The impact-induced tension as given by Eq. (11) depends on velocity and position. Thus to solve for these quantities it is useful to rewrite Eq. (1) in terms of these variables. This is easily done by using that $d^2z/dt^2 = dV/dt = V dV/dz$, where $z$ is the height of the chain above the table and $V = dz/dt$ is the velocity. Then Eq. (1) can be written as

$$V\frac{dV}{dz} = -g - \frac{\alpha V^2}{M + \gamma s \lambda}\frac{dM}{dz} \quad \text{(S13)}$$

where it has been used that $\lambda = dM/dz$. For a chain with constant density, $M = \lambda z$ and the first and last terms can be combined by multiplying the equation by the integration factor $2[M+\gamma s\lambda]^{2\alpha}$. Then Eq. (S13) can be written as

$$\frac{d}{dz}\{V^2[M + \gamma s\lambda]^{2\alpha}\} = -2g[M + \gamma s\lambda]^{2\alpha} \quad \text{(S14)}$$

This equation can be integrated over the fall of the chain, from $z = z_0$ (the initial length of the chain above the surface) to $z$. This yields Eq. (14) for the speed-squared of the chain as a function of height.

The experiments presented here measured both the speed and also $\Delta z =$ the separation distance between the two chains, when $z = 0$. This latter quantity can be calculated by using that $V = dz/dt$ and integrating to find the time the chain striking the surface takes to fall all the way to the surface

$$t = \int_0^{z_0} \frac{dz}{|V(z)|} \quad \text{(S15)}$$

This time can then be substituted into the equation of motion for the chain in free fall to find its height above the surface at time $t$, which is $\Delta z$.

$$\Delta z = z_0 + V(z_0)t - \frac{1}{2}gt^2 \tag{S16}$$

The solution so obtained is rather complicated because it depends on hypergeometric functions and gamma functions see Ref. (26). The solution can be simplified by making some approximations to get expressions that are much easier to use and still quite accurate.

In particular, taking $h = 0$ and working to just leading order in $\alpha$ yields Eq. (15) in the text. Another relatively simple analytical expression for $\Delta z/z_0$ exists for the limit $h \gg z_0$, which corresponds to when the speed of the chain is large as it first strikes the surface. This condition does not match those of the experiments presented here, but it is instructive for determining how large $\Delta z$ can be. In this limit the terms containing $g$ in Eqs. (S13) and (S16) are negligible, the integral in Eq. (S15) is easy to do, and the result is

$$\frac{\Delta z}{z_0} \approx \left(\frac{\alpha}{1+\alpha}\right) - \left(\frac{1}{1+\alpha}\right)\left(\frac{\gamma s}{z_0}\right)\left\{1 - \left[\frac{\gamma s}{z_0 + \gamma s}\right]^\alpha\right\} \tag{S17}$$

This expression is valid to all orders in $\alpha$. In the limit $\gamma s/z_0 \to \infty$ then $\Delta z/z_0 \to 0$, in agreement with the discussion in section III.B. In the limit of $\gamma s/z_0 \to 0$ this equation reduces to just the first term, and agrees with the numerical results presented in Ref. 26 for large initial speeds. It is somewhat surprising that the dependence on $h$ cancels out so that $\Delta z/z_0$ depends only on the chain parameters, just like for Eq. (15).

An indication of how the initial speed affects $\Delta z$ can be obtained by comparing Eqs. (15) and (S17) for the typical case of $\gamma s/z_0 \ll 1$. Numerically evaluating these equations shows that $\Delta z/z_0$ is only 20% larger for the case of very large initial speeds. Thus the effects of the impact-induced tension are only increased by a relatively small amount when the initial drop height is increased for a chain of fixed length.

The analytical expressions given here are only valid for a uniform density chain. For a chain with a variable mass density, the integrations generally cannot be performed exactly.

**S V. Lift-off from a horizontal surface.**

The previous sections have discussed the case of a rod being laid-down onto a surface. For completeness, here is discussed the lift-off of a rod from a horizontal surface. Pulling the chain upwards produces an upward force on the end of the chain. This motion enhancing force is responsible for the phenomenon known as the "chain fountain."

For the type of chain discussed here, one made of rods that are slightly tilted from horizontal and connected together by string at the ends, the lift-off of a rod from the surface is a two-step process. As the chain moves upwards, first the string on one side of the rod becomes taut and lifts that end of the rod off of the surface. This side will be denoted as the right side, to match the sketch of the forces shown in Fig. 5 (with the simplification that the surface is horizontal, $\theta = 0$, so the friction force $f$ is negligible).
The tension $F$ in the taut string causes the rod to try to rotate, pushing the left side of the
rod into the surface, and therefore causes the surface to push upwards on the left end of
the rod with the force $N$. This latter force is the force that enhances the upward motion of
the chain. The second stage of lift off is when the string on the left side of the rod
becomes taut and lifts that side of the rod off of the surface, causing the force $N$ to vanish.
To calculate the size of the upward force, $N$, we only need to model the first stage of the
lift-off process. Newton's laws of motion have already been applied to the rod in this
configuration and are given in Eqs. (3) and (4) in section III.A.

The discussion in section III.A and here differ in how the rod ends move during the
interaction with the surface. For lifting a rod off of the surface, the change in velocity of
the rod ends during the first stage of the lift-off process is parameterized here as

| | | |
|---|---|---|
| Left end, z-dir.: | $0 \rightarrow 0$ | (S18) |
| Right end, z-dir.: | $0 \rightarrow (1+\beta')V$ | (S19) |

Initially the bar is assumed to be stationary on the surface. Eqs. (S18) assumes that the
left end of the rod is in contact with the surface during the first stage of lift-off, and so its
vertical velocity does not change. Eq. (S19) parameterizes the interaction between the
chain and the right end of the rod as the string there become taut and this end of the rod
leaves the surface. This interaction can be thought of as a collision between the chain
and the rod (mediated by the connecting string), and so a coefficient of restitution, $\beta'$, has
been introduced to describe this interaction. $\beta'$ depends on the material properties of the
string that connects the rod and the chain. If the string damps all of the relative motion
between the chain and the rod, then the collision is completely inelastic and $\beta' = 0$.
However if the rod-chain interaction is not completely inelastic, then $\beta' > 0$. Eq. (S19)
also assumes that the chain above is long, and does not alter its motion from the force $F$
between it and the rod.

From Eqs. (S18) and (S19) it is straightforward to derive that during the first stage of lift-
off, the change in the center of mass velocity of the rod is $\Delta V_z = (1+\beta')V/2$ and the
change in the rotational velocity about the center of mass is $\Delta \omega = (1+\beta')V/L$. Using these
values, and Eqs. (3) and (4), the tension force acting on the rod, $F$, can be eliminated
from the calculation, giving an expression for $\int N dt =$ the impulse acting on the bottom
rod by the surface. This impulse can then be turned into an expression for the average
force by dividing the impulse by the time between collisions of successive rods with the
surface $= s/V$, where $s =$ the distance between the rods. Using that $m/s = \lambda =$ the linear
mass density gives for the average force $N = \alpha_{\text{lift-off}} \lambda V^2$, with

$$\alpha_{\text{lift-off}} = \left(\frac{1+\beta'}{4}\right)[1-\xi] \tag{S20}$$

The form for the motion enhancing force $N$ is the same as given in Eq. (2), as it has to be
from the dimensional analysis argument given in the introduction. The form for $\alpha$ given
in Eq. (S20) for lift-off is identical to that for lay-down given in Eq. (8). This similarity
is not a coincidence. Both situations are dynamically the same process; a force applied at
one end of the rod causes a "reaction" force at the other end of the rod. However the two
situations differ in that they depend on different material properties. Lay-down of a rod depends on the coefficient of restitution for the rod-surface collision, while lift-off depends on the coefficient of restitution for the rod-string-chain collision.

For uniform bars $\xi = 1/3$, and for a completely inelastic collision $\beta'=0$, then $\alpha_{\text{lift-off}} = 0.17$. This is somewhat similar in size to the values $\alpha_{\text{lift-off}} \approx 0.10\text{-}0.12$ found in observations of the chain fountain using bead chains. This value for bead chains is obtained from the experiments by using that $\alpha_{\text{lift-off}} \approx h_{\text{above}}/(h_{\text{above}}+h_{\text{below}})$ where $h_{\text{above}}$ is the distance the chain extends above the initial pile and $h_{\text{below}}$ is the distance from the initial pile to the floor. See Refs. 18 and 21 for measurements.